\documentclass[10pt, conference]{FMFP2022}
\usepackage[T1]{fontenc}
\usepackage[top=1.50cm, bottom=1.50cm, left=1.884cm, right=1.884cm, includehead, includefoot, heightrounded]{geometry}
\usepackage{fancyhdr} 
\usepackage{graphicx} 
\usepackage[textfont=bf, font = normalsize]{caption} 
\usepackage{floatrow} 
\usepackage{kantlipsum}
\usepackage{amssymb}
\usepackage{rotating}
\usepackage{multicol}
\usepackage{amsfonts}
\usepackage{amsmath}
\floatstyle{plaintop} 
\restylefloat{table} 
\usepackage{subfigure} 
\renewcommand{\headrulewidth}{0pt}

\begin{document}

\title{\LARGE{ \bf Performance enhancement of a 100 watts class Tesla turbine}}
\author{\textbf{Arindam Mandal\textsuperscript{1}, Rajosik Adak\textsuperscript{1}, and Sandeep Saha\textsuperscript{1}}
\\\textsuperscript{1}\small{Department of Aerospace Engineering, IIT Kharagpur, Kharagpur 721302, India}}

\maketitle
\thispagestyle{fancy} 
\pagestyle{plain} 

\noindent \textbf{ABSTRACT}
Tesla turbines are an attractive but less explored area in low-power applications. This article presents an experimental investigation of a centimeter scale Tesla turbine in bi and uni-directional outlet configuration with compressed air at 6 bar. The turbine's performance is enhanced by $\approx$ 38\% for a uni-directional outlet configuration. Furthermore, we also investigate the electrical power of the turbine in a bi-directional outlet configuration by coupling the turbine with a generator. Despite achieving higher performance in uni-directional outlet configuration, we observe substantial losses at the inlet we use for the experiment. To illustrate and improve the losses, we numerically investigate the turbine inlet at a total pressure and temperature difference of 2 bar and 50$^\circ$C, respectively. Subsequently, we design two more nozzles and compare their performance with the nozzle we used in our experiment. Our findings suggest that nozzle 3 performs the best in delivering the highest Mach no and uniformity across the slits. This observation would help optimize the nozzle suitable for the Tesla turbine.

\section{{\textbf{INTRODUCTION}}}
\begin{figure}[!h]
\centering
\fbox{\includegraphics[width=0.75\textwidth]{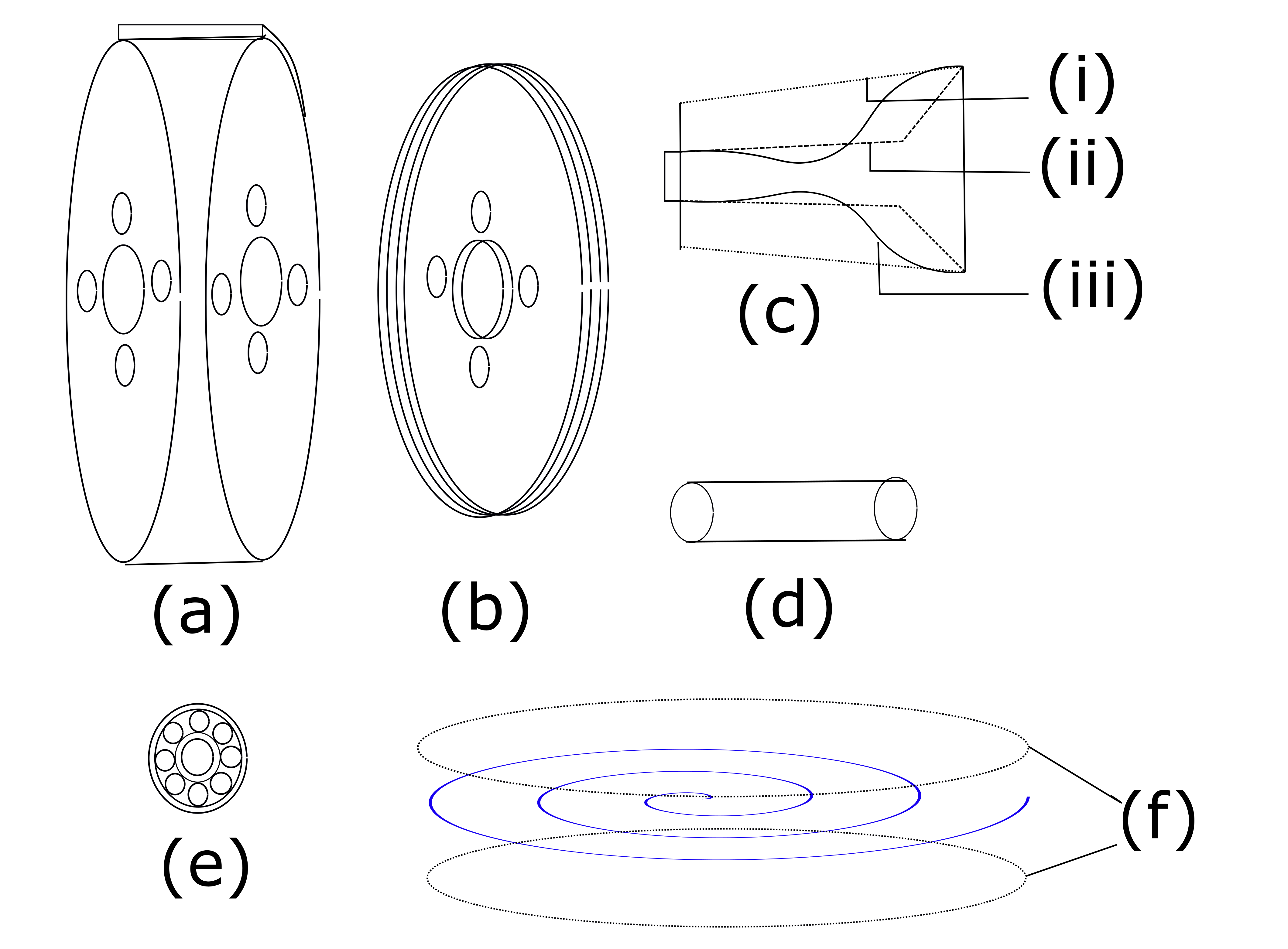}}
\caption{Schematic of the exploded view of the turbine. a: Casing, b: Rotors, c: Inlet configurations (i,ii and iii), d: Shaft, e. Bearing. (f) shows the fluid flow between two consecutive corotating discs}
\label{conference_schematic}
\end{figure}

Small scale low microturbines are crucial since there is a growing need for them in numerous applications. Examples of notable applications include waste heat recovery, pico-hydro power, organic Rankine cycle technologies, biomass, waste heat recovery, micro GT, and many more. The increasing need for energy harvesting at these scales poses a variety of challenges to the performance and manufacturability of the turbine due to their compact sizes, high rotational speeds, and increased viscous losses. An alternative expansion device addressing these challenges can be highly beneficial regarding its techno-economic feasibility. The tesla turbine is one of its kind, which has a uniquely simple design and a unique mechanism of momentum transfer. The turbine rotor consists of several co-axial discs closely packed with each other. Each of the rotor discs has outlet ports near the center. A casing covering the rotors helps guide the flow from the inlet. The turbine shaft is attached to the rotor-casing configuration with the help of two bearings. Figure \ref{conference_schematic} shows the different components of the turbine. After being injected through the inlet system of the turbine, the fluid flows spirally inward and exits through the ports located near the shaft in the axial direction. The momentum transfer from the fluid to the rotor takes place using the adhesion and viscosity of the fluid. This mechanism makes the turbine attractive at small scales where the dominance of the viscous force becomes significant.

This turbine was conceptualized by Nicola Tesla \cite{tesla1913} in the year 1907. Initially, the device was unable to attract market attention because of the no potential requirement of low-power harvesting technologies. After that, till the twentieth century, a considerable amount of thrust was given to the possible design modifications \cite{oklejas1975gas, oklejas1976tesla, harold2002tesla}, improved seal designs \cite{caldwell1973efficiency} and loss analysis \cite{crawford1974,rice2003tesla} of the turbine and its ancillary components for enhancing performance. In addition, a few simplified theoretical models were developed \cite{armstrong1952,lawn1972investigation} to get an insight into the influence of the parameters associated to the turbine. 

More recently, there has been a surge in interest in exploring this technology due to its several advantages over conventional energy harvesting technologies. Due to its simple design and flow mechanics, the turbine can handle particle-laden fluids and two-phase expansion with minimal damage. Numerical simulations of the gas flow and mass transfer between two coaxially co-rotating discs were performed by Sandilya et al. \cite{sandilya2001numerical} where they found a satisfactory match between the numerical and experimental value of mass transfer coefficient. Using numerical simulations, Ladino \cite{ladino2004numerical} presented the load coefficient curve, efficiency, and degree of reaction variation after maintaining a constant rotational speed. Upon developing an analytical model, Deam et al. \cite{deam2008} scaled down the turbine in millimeter size to achieve better efficiency. Lemma et al. \cite{lemma2008} investigated the Parasitic, viscous, and other dissipative losses in the bearings and end walls to mitigate their effect on decaying the performance. \\

An explicit description of a flexible test rig was developed by Hoya et al. \cite{hoya2009} where they calculated the low torque at high RPM using the angular acceleration method. A comparative study in the efficiency offered by Tesla and small bladed microturbine in a micro power plant was addressed by Lampart et al. \cite{lampart2009design,lampart2011investigations} to establish the competitiveness of a Tesla turbine. To understand the transport phenomena inside the turbine rotor, a number of numerical and analytical models solving the Navier-Stokes equations using different approaches are present in the literature repository \cite{kavenuke2009modeling,guha2013,sengupta2012theory,guha2014similitude,romanin2012,schosser2015,schosser2017analytical,ciappi2019computational}. Aside from that, flow diagnostics using particle tracking velocimetry by Schosser et al. \cite{schosser2016three} provides an insight into the component-wise velocity profiles inside the rotor gaps at different radial locations. In recent years,  A wide range of application based studies of Tesla turbine related to Micro-air vehicles \cite{mandal2017performance}, Organic Rankine cycle \cite{talluri2018design,talluri2020experimental,dumont2019comparison,manfrida2017revised,manfrida2018upgraded}, Combined heat plant \cite{carey2010}, Pico-hydro applications \cite{ho2011, choon2011optimization, krishnan2011micro, krishnan2015}, Ammonia synthesys \cite{han2021self} have been investigated or under investigation \cite{niknam2021numerical,pacini2020computational}.

The recent resurgence in interest indicates the importance of investigating the turbine further to mitigate the losses due to the nonuniformity and disturbances associated with the inflow to rotor and rotor to outflow interaction. This article presents the experimental investigation of a centimeter-scaled Tesla turbine in uni and bi-directional outlet configuration with compressed air. We compare the performance in terms of Mechanical power output for the two configurations. Furthermore, we investigate the losses in the inlet due to the sharp divergent and the slit configuration at the inlet rotor junction. Finally, we compare the nozzles' performance by looking at the peak discharge Mach number and channel-wise disparity in flow injection to the rotor. Our numerical results can be beneficial in coming up with better inlet configurations for extracting maximum energy from the fluid, which leaves a further scope for further investigation.
 
 \section{\textbf{METHODOLOGY AND EXPERIMENT}}\label{sec2}
\begin{figure}[!h]
    \centering
    \includegraphics[width=0.46\textwidth]{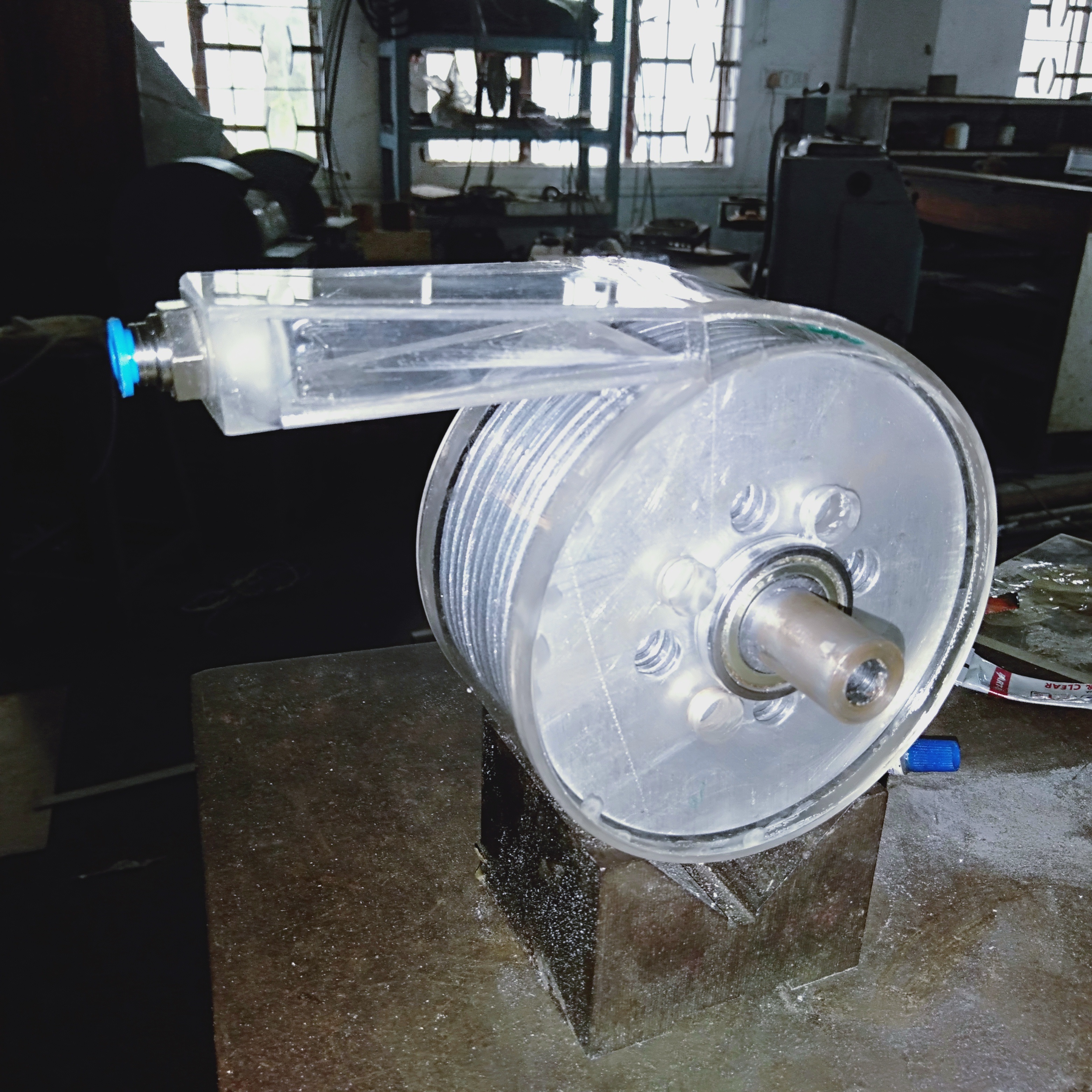}
    \includegraphics[width=0.43\textwidth]{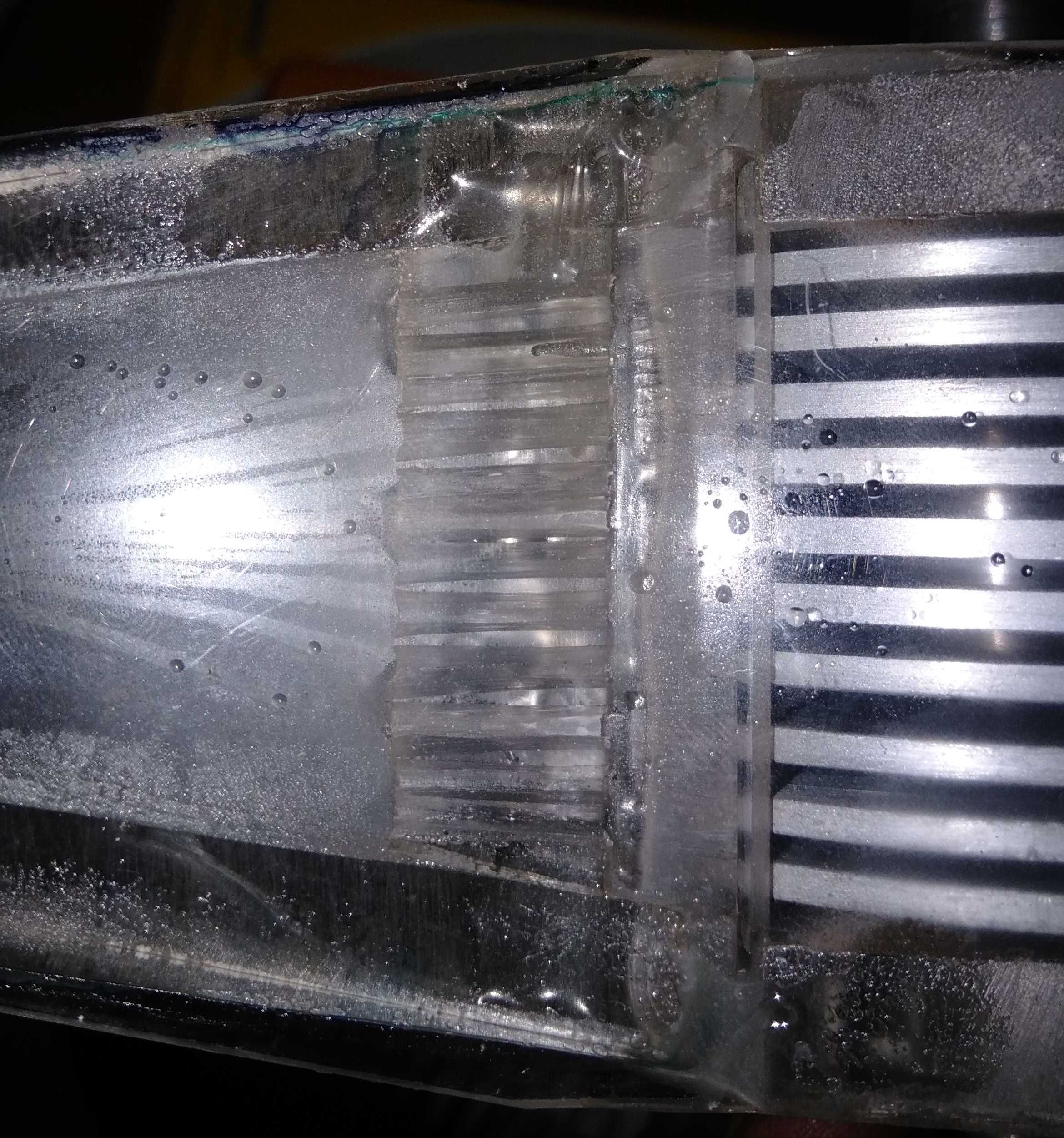}\\
    (a) \hspace{2 in} (b)\\
    \caption{(a) Fabricated turbine and (b) the inlet system}
    \label{fabricated_turbine}
\end{figure}

 A lab scale turbine prototype (in fig. \ref{fabricated_turbine}a) is fabricated for experimentation in the Aerospace Engineering department, IIT Kharagpur. The turbine consists of 10 consecutive corotating discs having an outer diameter of 10 cm and a thickness of 2 mm. The gap between the consecutive rotating discs is 2 mm, and the distance between two extreme discs to the adjacent casing wall is 1 mm. The turbine has four outlet holes near the shaft, having a center distance of 2 cm from the shaft center. The shaft and the exhaust ports' diameters are 1.5 cm and 1 cm, respectively. The distance between the shroud and the disc's edge is 1 mm.
The inlet system of the turbine is designed to connect the compressor outlet port of 6mm dia to the turbine having an inlet of 4 cm thickness. The area of the inlet is 0.4 $cm^2$, which is segregated using slit configurations to guide air to each gap with minimum interaction with the peripheral walls of the rotor.

\begin{figure}[!h]
\centering
\includegraphics[width=1\textwidth]{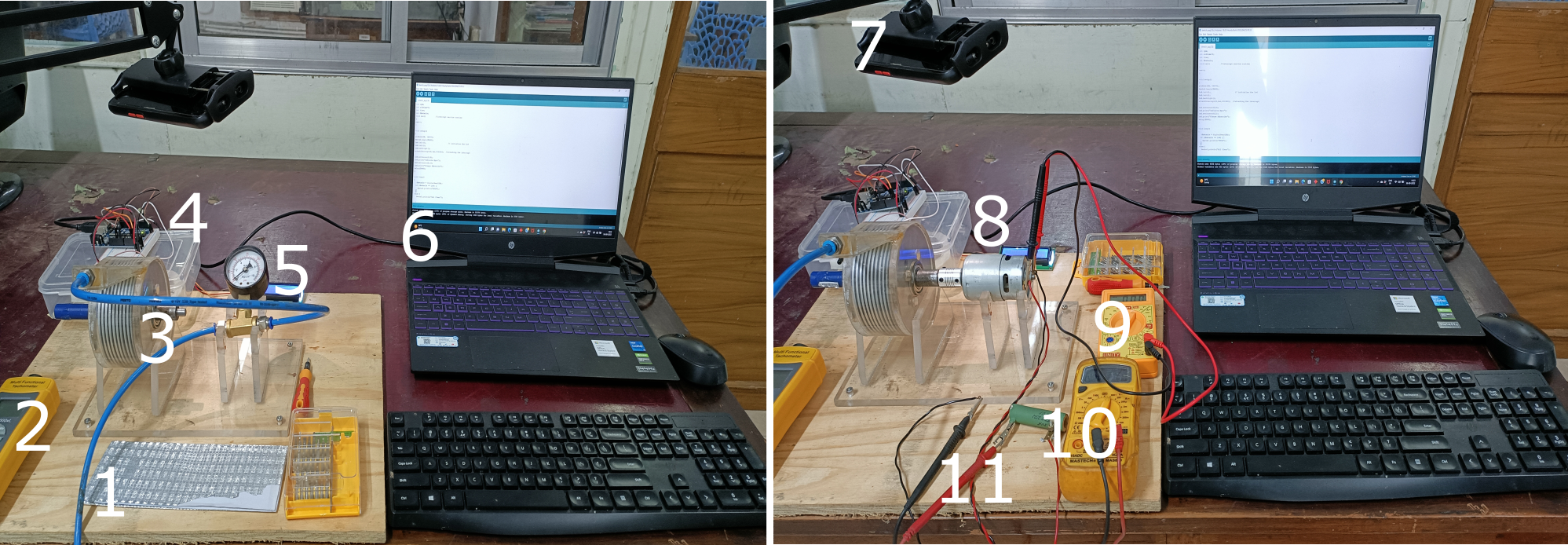}\\
(a) \hspace{1 in} (b)
\caption{Details of the experimental set up. 1- Compressor discharge, 2- Tachometer, 3- Turbine, 4- Arduino based Tachometer, 5- Pressure gauge, 6- PC for data acquisition, 7- Camera, 8- Generator, 9- Multimeters, 10- Rheostat, 11- Electrical circuit}
\label{experimental_setup}
\end{figure}

The turbine inlet is connected to the compressor by Polyeurathane pneumatic pipes with an installed pressure gauge. The RPM of the turbine is measured using an Arduino-based tachometer. The angular acceleration-deceleration approach computes the net accelerating and decelerating torque.The experiment also uses a digital tachometer to validate the turbine's RPM. We conduct our experiment at 6 bar of inlet pressure for uni-and bi-directional outlet configuration. The detail of the experimental setup can be seen in the figure \ref{experimental_setup}.

\section{\textbf{RESULTS AND DISCUSSION}}\label{sec3}
\subsection{\textbf{Performance characteristics}}\label{sec3p1}
\begin{figure}[!h]
    \centering
    \includegraphics[width=0.44\textwidth]{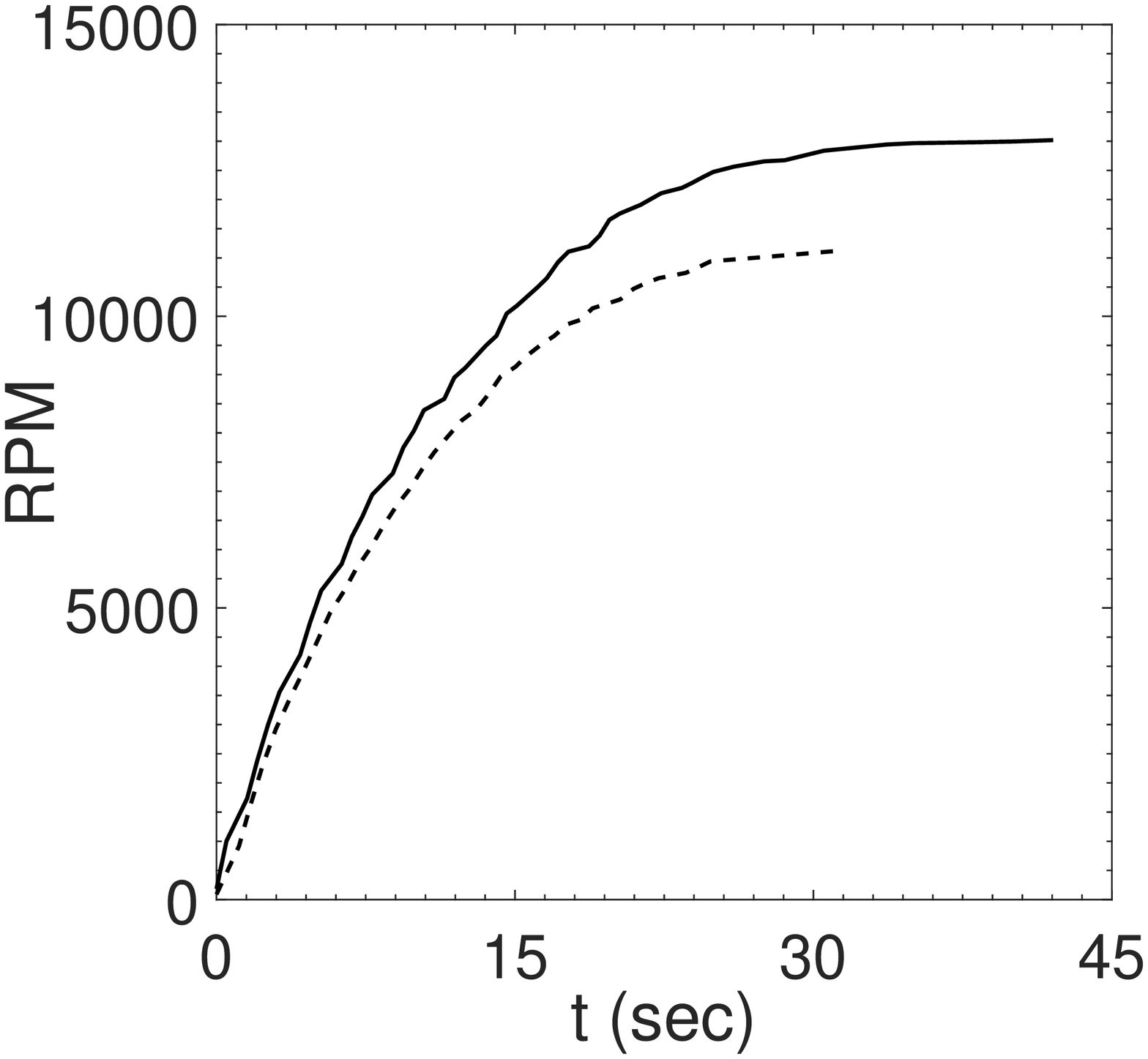}
    \includegraphics[width=0.45\textwidth]{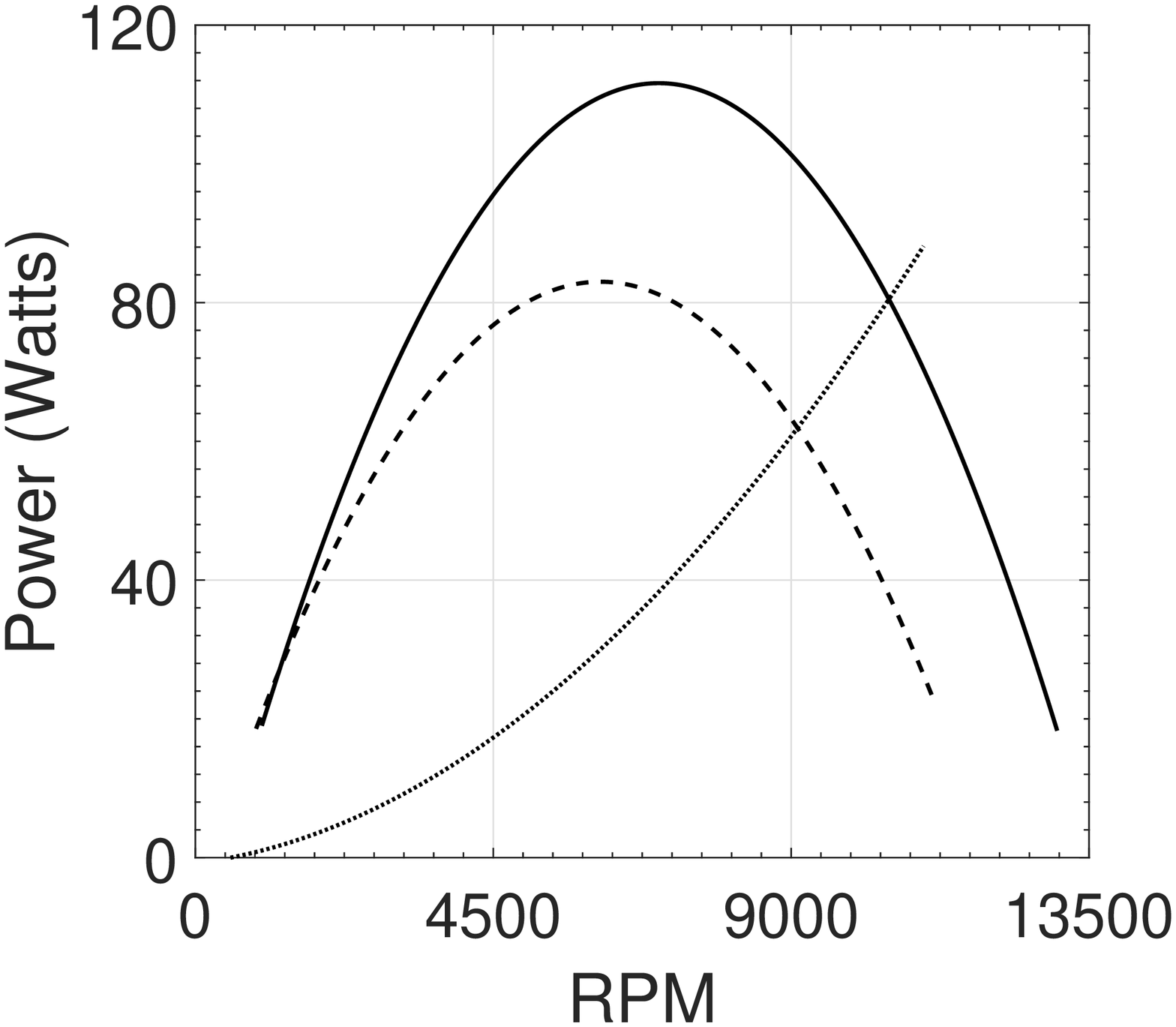}\\
    (a) \hspace{ 1.25 in} (b)
    \caption{Distribution of (a) RPM with t; (b) Power due to accelerating torque with RPM for bi-directional (dashed line) and uni-directional (solid line) outlet. Dotted line represents the power due to braking torque.}
    \label{Bi_directional_Power_vs_RPM}
\end{figure}

We tabulate the RPM with time with the help of a serial monitor and calculate the angular acceleration as a function of RPM. Once the turbine reaches its stable RPM, we shut off the compressed air sypply and continue to perform data acquisition until the turbine becomes stationary. We continue the similar process for a uni-directional outlet case. We observe from figure \ref{Bi_directional_Power_vs_RPM} that the turbine with a uni-directional outlet accelerates faster than the turbine with a bi-directional outlet configuration. In addition, for a supply pressure of 6 bar, we achieve a maximum RPM of 13019 and 11124 for uni and bi-directional outlet configurations, respectively. Figure \ref{Bi_directional_Power_vs_RPM} shows the power variation due to accelerating and braking torque. There is an increment of $\approx$ 38\% in power due to accelerating torque for a uni-directional outlet configuration.

\begin{figure}[!h]
    \centering
    \includegraphics[width=0.65\textwidth]{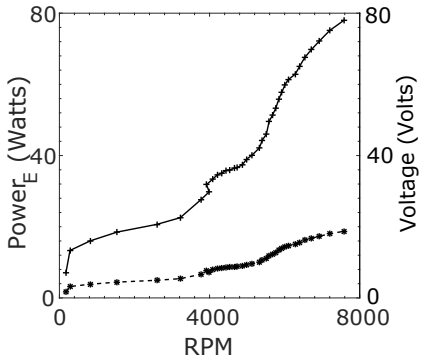}
    \caption{Experimental results of voltage (dashed line) and electrical power (solid line) at different RPM}
    \label{experimental_result}
\end{figure}

We integrate the turbine with a 100 W class Fedus RS-775 DC electric motor to measure the electrical power output. However, the motor is used as a generator to measure the output voltage and current. The electrical circuit is attached to a rheostat, and the experiment is performed with the 5-ohm load resistance. The figure \ref{experimental_result} illustrates the motor's voltage and power variation at various RPM. The turbine-generator can produce a maximum of 78 watts of electrical power at 7800 RPM for bi-directional outlet configuration.

\subsection{\textbf{Inlet design}}\label{sec3p2}
\begin{figure}
    \centering
    \fbox{\includegraphics[width=0.8\textwidth]{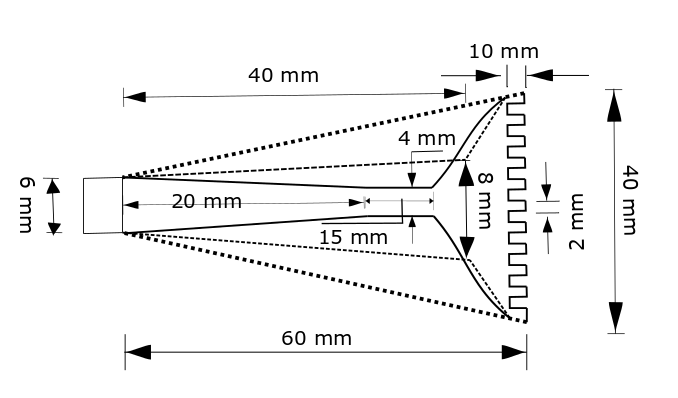}}
    \caption{Schematic diagram of the nozzles. The dotted, dashed and solid lines represent nozzle 1, nozzle 2 and nozzle 3, respectively.}
    \label{nozzle_schematic}
\end{figure}

Designing the inlet is the most crucial part of the turbine where the maximum loss occurs. Notably, the compressor and back pressure at the inlet rotor connection point regulates the flow across the nozzle. In the present article, we only consider the inlet section for analysis. The details of the nozzle dimensions are in figure \ref{nozzle_schematic}.
\subsubsection{Numerical methodology and grid Grid sensitivity}
\begin{figure}[!h]
   \centering
    \begin{turn}{-90}\fbox{\includegraphics[width=0.43\textwidth]{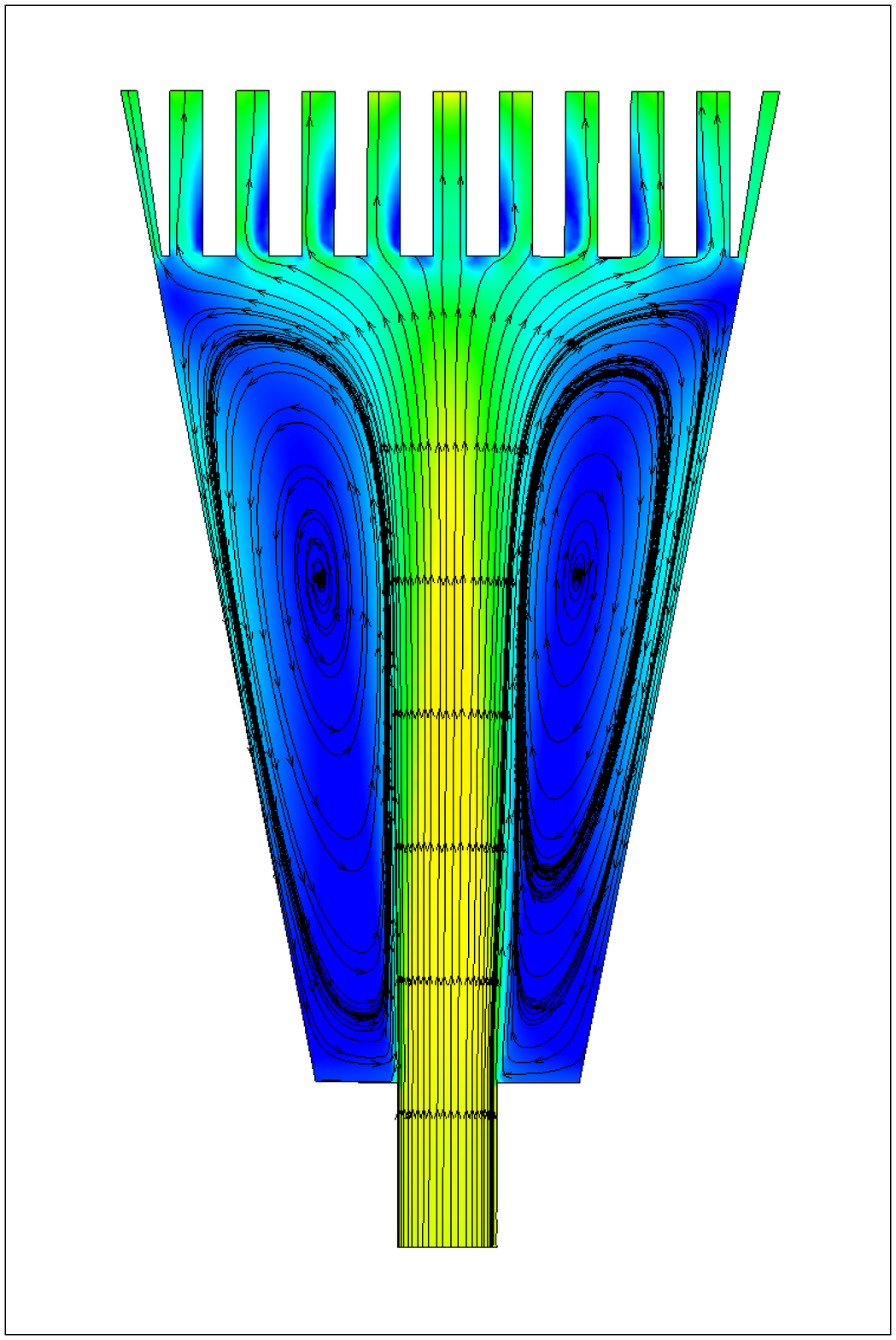}}\end{turn}\\
    \vspace{2 mm}
    (a)\\
    \begin{turn}{-90}\fbox{\includegraphics[width=0.43\textwidth]{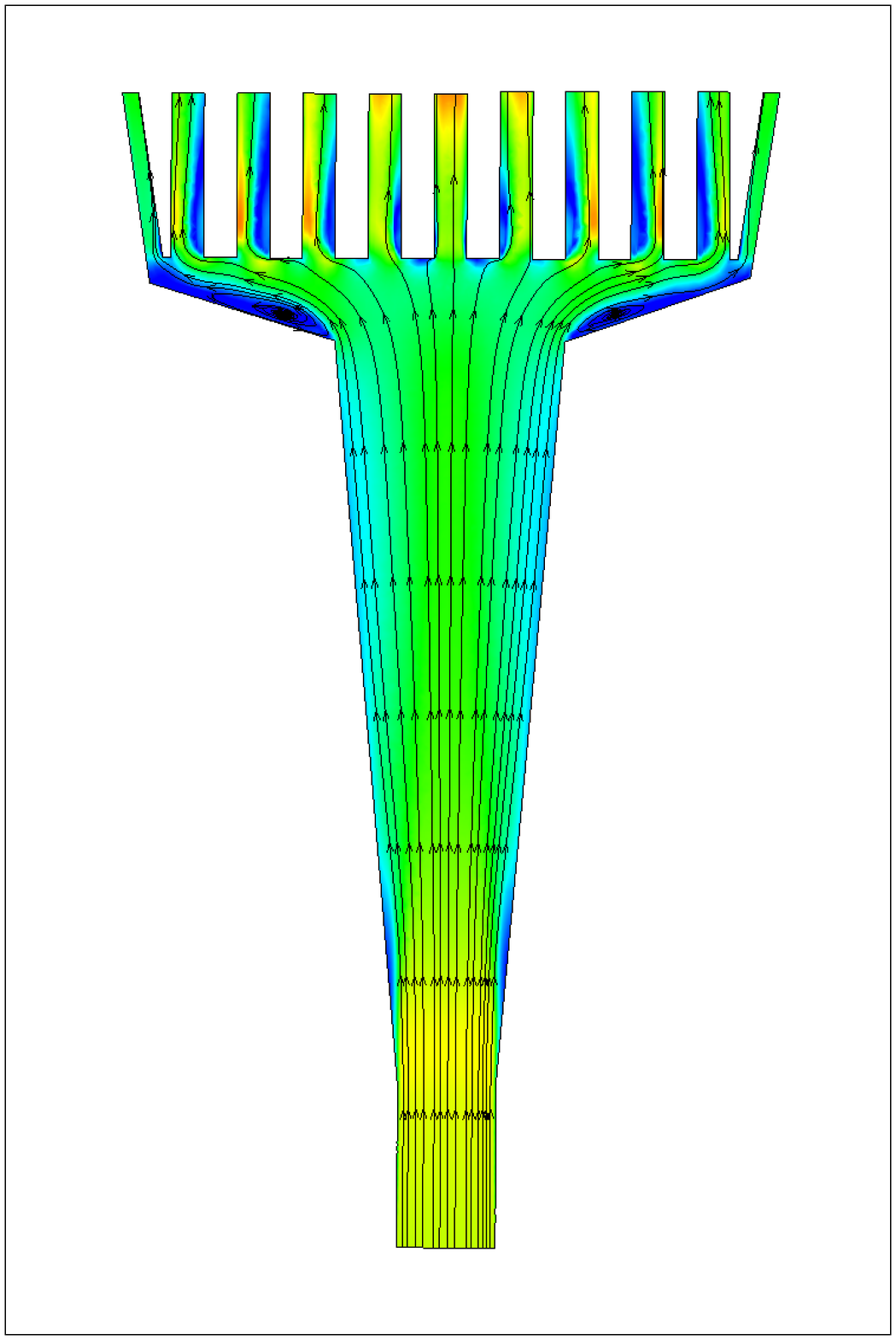}}\end{turn}\\
    \vspace{2 mm}
    (b)\\
    \begin{turn}{-90}\fbox{\includegraphics[width=0.43\textwidth]{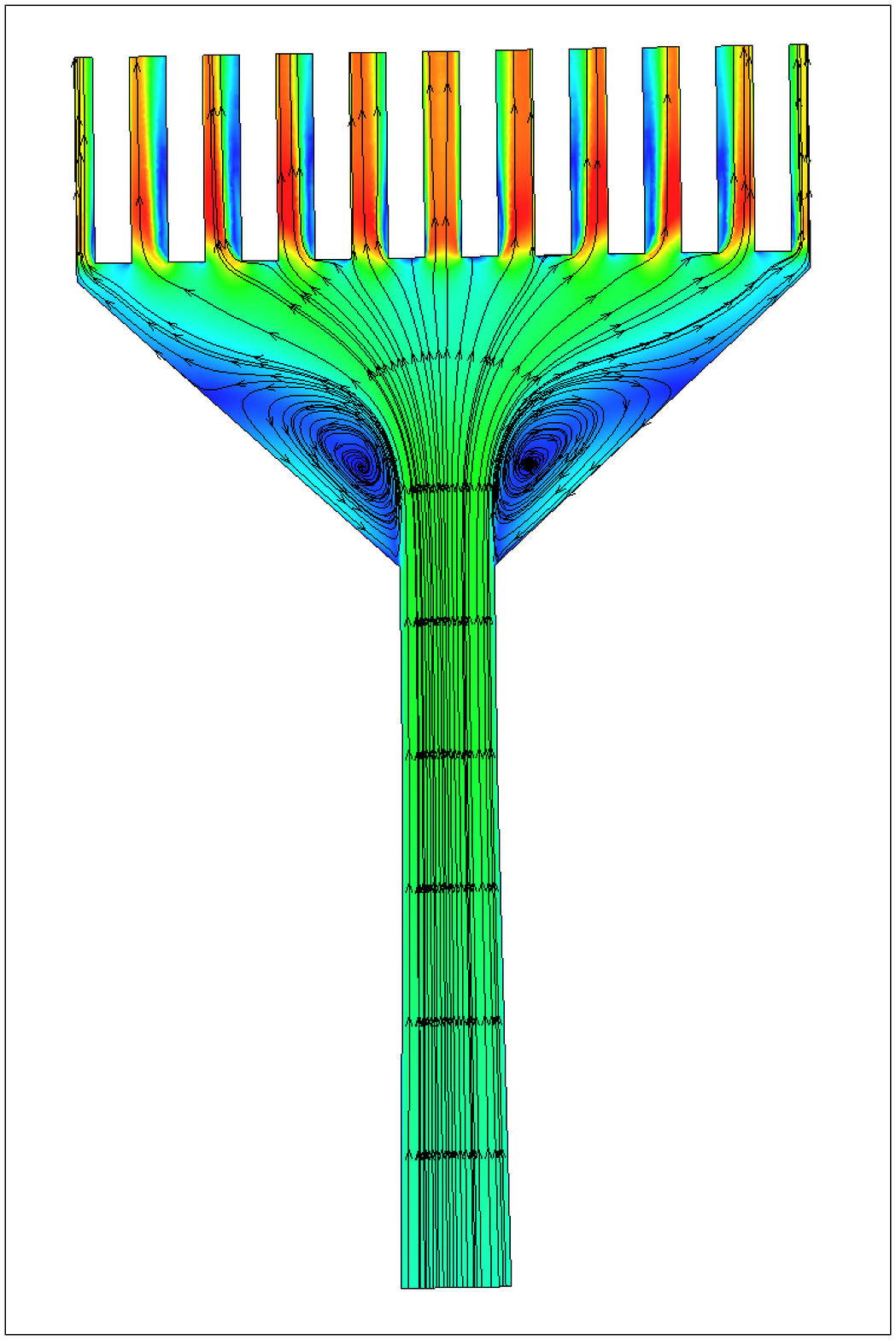}}\end{turn}\\
    \vspace{2 mm}
    (c)\\
    \vspace{2 mm}
    \includegraphics[width=0.8\textwidth]{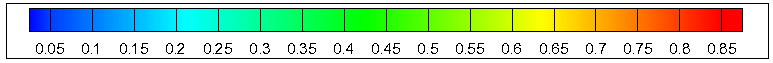}\\
    \caption{ Distribution of the Mach number for (a) nozzle 1 (b) nozzle 2 (c) nozzle 3.}
    \label{Mach_streamline}
\end{figure}

\begin{figure}[!h]
    \centering
    \includegraphics[width=0.65\textwidth]{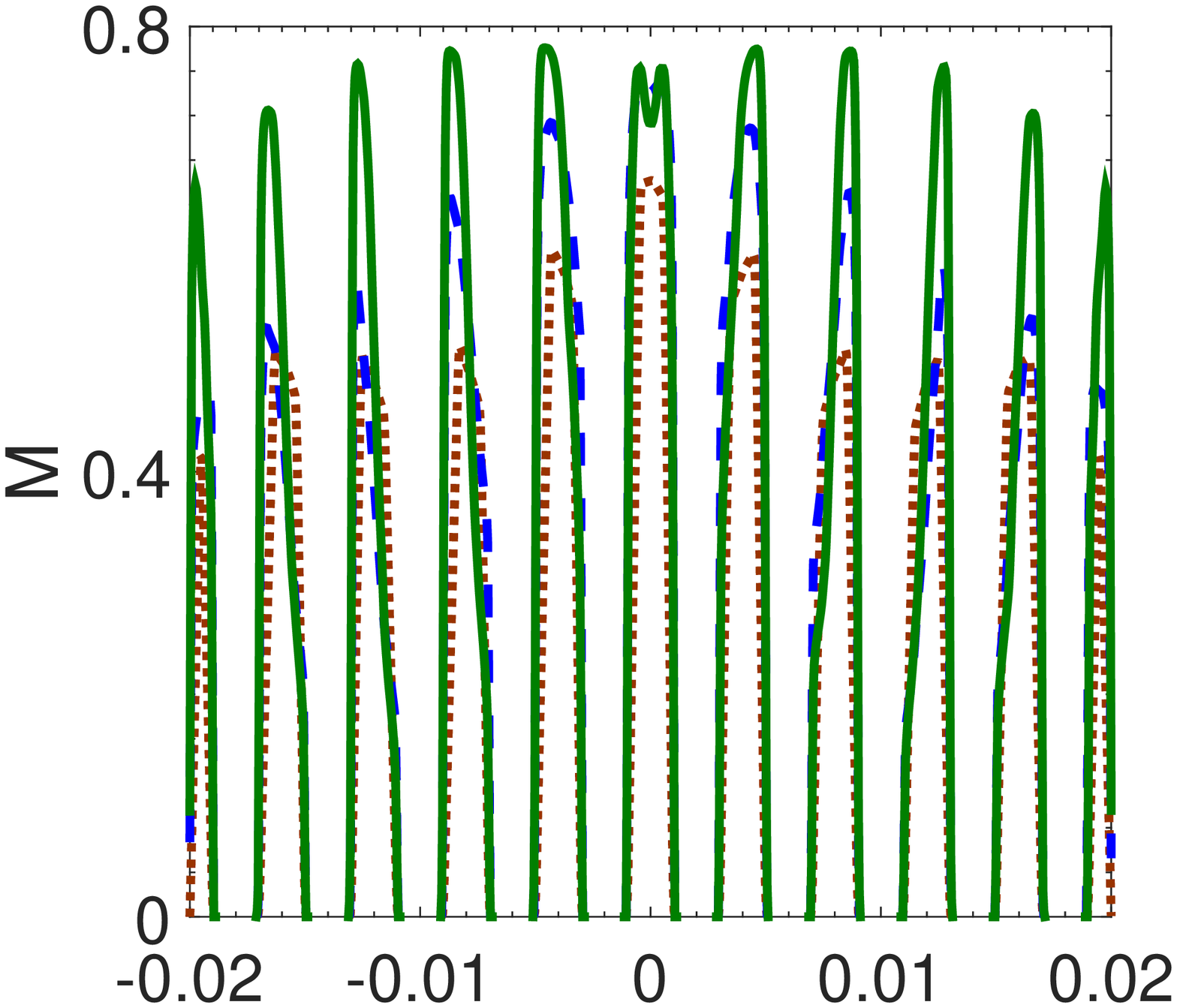}\\
    (a)\\
    \vspace{0.1 in}
    \includegraphics[width=0.32\textwidth]{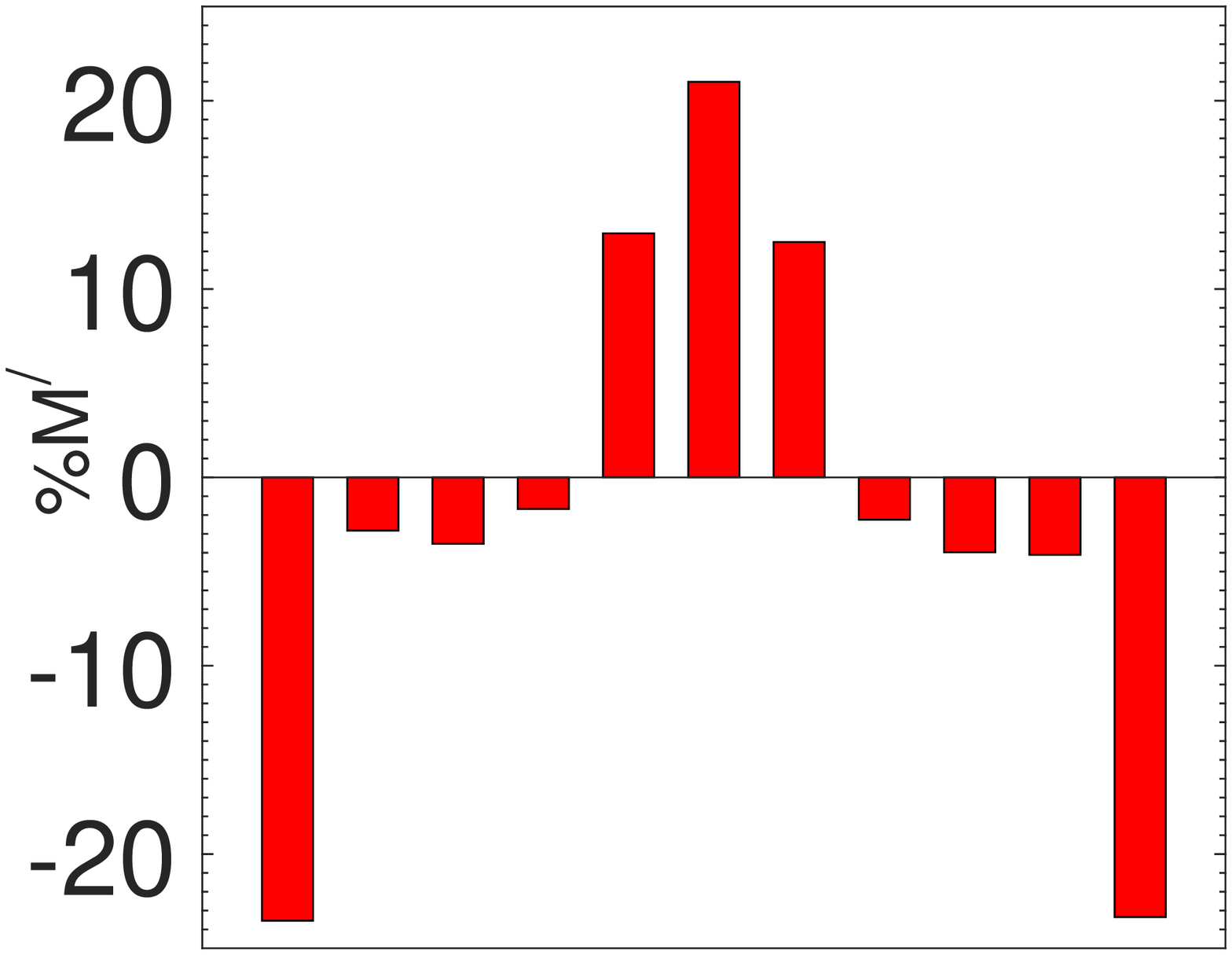}
    \includegraphics[width=0.32\textwidth]{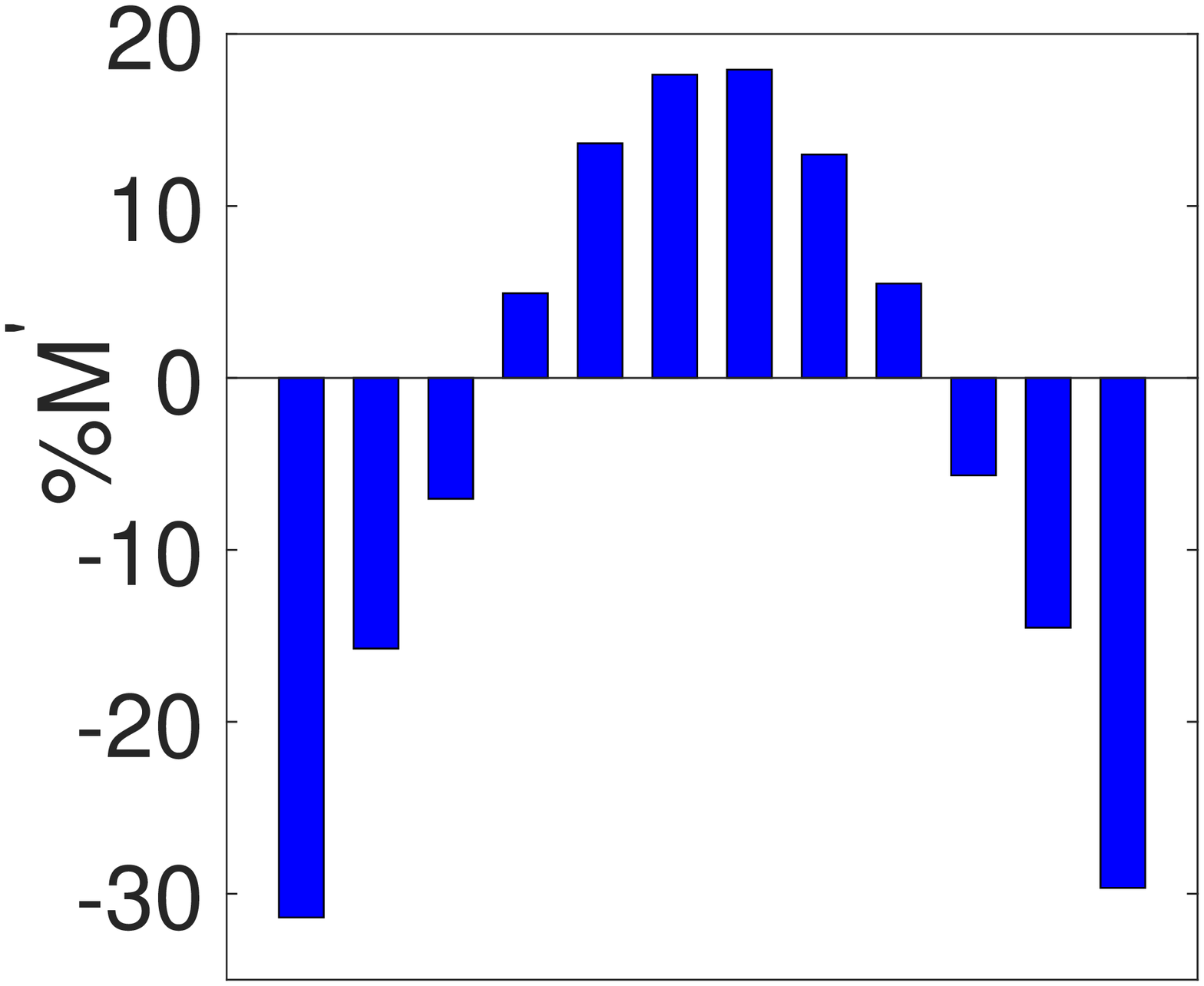}
   \includegraphics[width=0.32\textwidth]{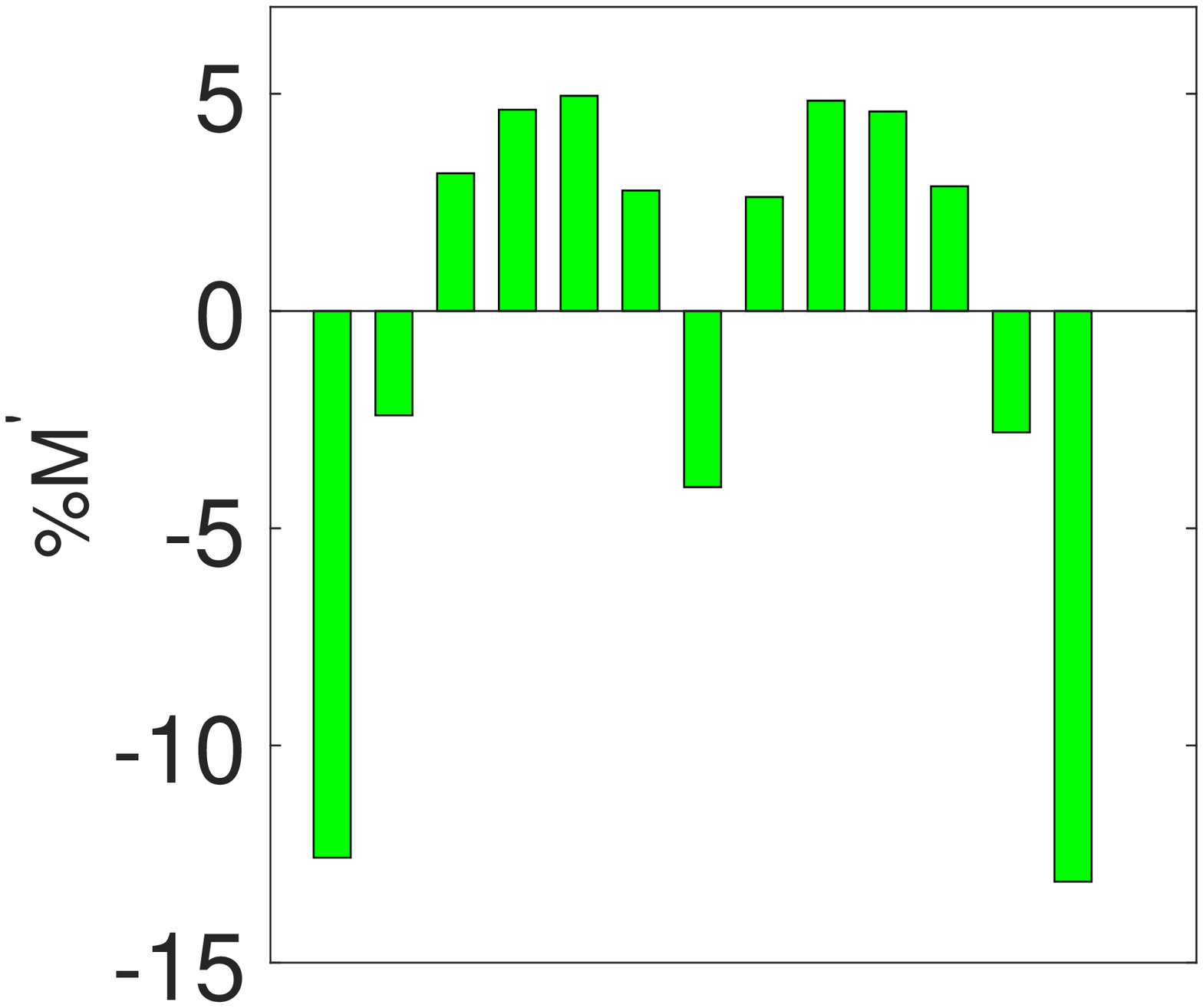}\\
   (b) \hspace{0.8 in} (c)\hspace{0.8 in} (d)
    \caption{ (a) Comparison of peak Mach number across the slits. The dotted, dashed and solid lines represent First, second and third nozzle, respectively. (b, c, d) The \% disparity in peak Mach no at discharge for three nozzles.}
    \label{Mach_deviation}
\end{figure}

The Inlet section of the nozzle is a pressure inlet boundary where the total pressure is at 6 bar. We consider the outlet of the nozzle is  at 4 bar. The temperature at the compressor discharge and the inlet-rotor junction are 450K and 400K, respectively. The surface of the nozzle is a no-slip type boundary. Considering these boundary conditions, we solve governing compressible Reynolds-averaged Navier-Stokes equations using the $K-\omega$ SST turbulence model in the commercial CFD package Fluent 2021 R2. The numerical domain is discretized using the body-fitted tetrahedral grids, maintaining a wall $y^+\approx O(1)$. The table \ref{grid_sensitivity} below presents the grid sensitivity study. As the desired output shows a variation $\leq 2\%$,  We conduct the subsequent numerical simulations using grids with $\approx$ 1M elements.
\begin{table}[h]
    \centering
    \begin{tabular}{c c c c c}
    \hline
      Grid type & Elements & \multicolumn{3}{c}{Outlet area averaged Mach no}\\
      \hline
         &  & Nozzle 1 & Nozzle 2 & Nozzle 3  \\
        \hline
        Coarse & $\approx$ 0.5 M & .5093 & .4993 & .5296 \\
        Medium & $\approx$ 1 M & .5089 & .4979 & .5289\\
        Fine & $\approx$ 2 M & .5080 & .4971 & .5284\\
        \hline
    \end{tabular}
    \caption{Grid sensitivity of the three inlet configurations}
    \label{grid_sensitivity}
\end{table}
The inlet design presented in this article is based on a two-pronged objective. (a) To Maximize the Mach number peak (b) To minimize the disparity in the Mach number peaks through every slit. To understand the loss mechanism and the flow behavior, we conduct a series of numerical simulations Where Nozzle 1 is the replica of the inlet nozzle considered for the experiment. Due to the abrupt divergent section and the presence of a large recirculation zone enclosed by a strong shear layer, we detect significant losses in Mach no. The dividing streamlines seen in figure  \ref{Mach_streamline} (a)  are considered when designing the following two inlet systems. The second nozzle we design is to eliminate the effect of recirculation. We gradually increase the inlet nozzle area until we reach the section where the flow bends in the previous observation. Despite the improvement in the average peak Mach number observed from figure \ref{Mach_streamline} (b), the disparity in the discharge Mach number through the slits increased. We consider designing the third nozzle as a converging-diverging type nozzle where we place the throat section at $x/L\approx 0.6$ to provide sufficient scope for flow to bend. Figure \ref{Mach_deviation} represents the peak discharge Mach number through the nozzles and the associated \% disparity from the mean peak Mach no through each slit. The comparison shows that the third nozzle offers maximum peak Mach number along with minimum \% deviation from the mean peak Mach no. It is necessary to note that the differential pressure between the upstream and downstream sections, the fluid, and the fluid's thermo-physical characteristics significantly influence the nozzle design. The nozzle's optimal size and shape might differ depending on these conditions.

\section{\textbf{CONCLUSIONS}}\label{sec4}
The article presents a preliminary experimental investigation of a centimeter scaled Tesla turbine using compressed air as a working medium and suggests an improved inlet design that could deliver higher achievable power compared to the inlet nozzle considered for this experiment. The key findings of the article are as follows;
\begin{enumerate}
    \item  Experimental investigation of the turbine conducted for uni and bi-directional outlet configuration at 6 bar inlet pressure.
    \item  uni-directional outlet configuration offered a peak power output of $\approx$ 110 watts at $\approx$ 7000 RPM. Whereas, bi-directional outlet configuration a peak output of 80 watts at 6500 RPM.
    \item  The turbine-generator configuration produced a peak electrical power output of 78 watts at 7800 RPM.
    \item  While assessing the inlet, we observe that nozzle 2 offers better peak Mach number than nozzle 1, but due to the losses accounted by the sharp bend, the disparity in \% deviation of the peak mach numbers from mean peak Mach no is nearly 30\%.
    \item  Nozzle 3 performs the best among all three nozzles. It offers a maximum peak Mach no with minimum \% deviation from mean peak Mach no is reduced to 12\%.
\end{enumerate}
The turbine's efficiency depends on several factors, e.g., RPM, disc gap, rotor radius, rotor to casing clearance, outlet configurations, fluid properties, and many more. Designing an efficient Tesla turbine could bring down the cost of a turbine substantially as compared to the other existing technologies because of its simplistic design. The above findings presented in this article could be helpful in the design improvement, thus leaving a plausible scope for further investigation.\\

 \noindent
\textbf{NOMENCLATURE}\\

\begin{tabular}{ccc}
$t$ & time & [sec]\\
$x/L$ & Non-dimensional-axial location & [--]\\
$k$ & Turbulent kinetic energy & [J/kg]  \\
$\omega$ & Specific dissipation rate & [1/sec]\\
$M$ & Mach no & --\\
$M'$ & \% Deviation from peak mean Mach no & --\\
\end{tabular}

\bibliographystyle{amsplain}
\bibliography{reference.bib}

\end{document}